\documentclass[conference]{IEEEtran}
\IEEEoverridecommandlockouts
\usepackage{cite}
\usepackage{amsmath,amssymb,amsfonts}
\usepackage{algorithmic}
\usepackage{graphicx}
\usepackage{academicons}
\usepackage{textcomp}
\usepackage{xcolor}
\def\BibTeX{{\rm B\kern-.05em{\sc i\kern-.025em b}\kern-.08em
    T\kern-.1667em\lower.7ex\hbox{E}\kern-.125emX}}
\begin{document}

\title{Capacity Higher-Order Statistics Analysis for $\kappa-\mu$ Fading Channels with Correlated Shadowing
%\thanks{978-1-6654-0306-1/21/\$31.00 ©2021 IEEE}
}
%
%\title{Capacity higher-order statistics analysis for $\kappa-\mu$ fading channels with correlated shadowing
%\thanks{978-1-5386-0585-1/17/\$31.00 ©2017 IEEE}
%}

%ORCID: 0000-0001-9308-4386 \href{https://orcid.org/0000-0003-4221-7622}
\author{\IEEEauthorblockN{Aleksey~S.~Gvozdarev,~\IEEEmembership{Member,~IEEE}}
\IEEEauthorblockA{\textit{Department of Infocommunications and radiophysics} \\
\textit{P.G.~Demidov Yaroslavl State University}\\
Yaroslavl, Russia \\
asg.rus@gmail.com, {\includegraphics[scale=0.1]{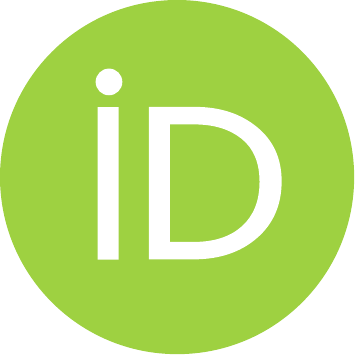}}~0000-0001-9308-4386}
%\and
%\and
%\and
%\and
%\and
}

\maketitle

\begin{abstract}
The proposed research performs an analysis of the capacity higher-order statistics for a single-input multiple-output multiantenna wireless communication system equipped with a maximum-ratio combining scheme. It was assumed that the propagation multipath channel is described with the $\kappa-\mu$ fading model with the correlated dominant components. Closed-form and asymptotic expressions were derived and applied to the problem of minimum capacity reliability (due to channel fluctuations, thus possible rate deterioration) and corresponding signal-to-noise ratio analysis. The performed computer simulation, verifying the correctness of the obtained expressions, along with the generalized $\kappa-\mu$ fading channel with correlated shadowing, assumed several specific limiting simplified cases: Rayleigh, Rician and Nakagami-$m$. It was shown that the signal-to-noise ratio (at which minimum capacity reliability is attained) is achieved at greater values than that of simplified models, and the absolute value of this minimum can be smaller/higher than for the degenerate cases depending on the dominant components one-step correlation coefficient.
\end{abstract}

\IEEEpeerreviewmaketitle

\section{Introduction}
\label{intro}
Diversity reception (SIMO, MISO, MIMO) \cite{Pau08}, \cite{Gol05} is one of the key technologies in wireless communications helping to achieve transmission rate and reliability required by up-to-date communication standards (5G \cite{Sha20} \cite{Agi16}, WiFi 7 \cite{Kho20}, \cite{Den20}, etc.). At the same time, the growth of consumer demands leads to the strengthening in signal processing requirements: the increase of the desired link quality level (generally described by the error rate) increases the impact of the factors that were not significant at lower levels \cite{Sim05} (i.e. channel model or system configuration parameters, etc.). The two main strategies, assumed in practice to cope with that problem, are $1)$ to revise the existing description of the system, in general, and channel model, in particular, complexifying them to include greater details of signal propagation, $2)$ to derive communication schemes subjected to a larger number of restrictions considering more factors affecting link quality.

Within the first approach, the dominant role pertains to so-called generalised models that account for effects of path loss, fading (fast and slow) and shadowing \cite{Yac07a}, \cite{Yac07b}, \cite{Par14}, \cite{Bha15}. They are usually constructed mathematically in such a way as to incorporate less general ones as specific limiting cases (including Rayleigh, Rician, Nakagami, Hoyt, Weibull, etc.). Among the most widely used, the prominent role is given to the correlated $\kappa-\mu$ shadowed fading channel model \cite{Bha15}.

Commenting on the second approach one should mention the channel capacity higher-order statistics (CHOS) methodology (very popular nowadays), which in addition to existing restrictions parameterises communication quality and reliability utilizing statistical moments of channel capacity fluctuations (and hence rate fluctuations) \cite{Sag09,Yil12,Yil14,Tsi16,Pep18}.

Although the combination of those approaches can help to get a better insight on how to increase link quality, their joint implication in many cases leads to analytically intractable description, which lessens possible benefits.

Motivated by the problem stated above, the proposed research derives the representation of the capacity higher-order statistics for the SIMO system, functioning in presence of a multipath fading channel subjected to the $\kappa-\mu$ model with correlated shadowing. It is demonstrated that the assumed goal can be achieved by representing the capacity statistics in terms of the moment generating function of the signal-to-noise ratio; thus, the expressions for the moment generating function for the SIMO system employing maximum-ratio-combining (MRC) scheme was derived. The derived solution was then analyzed and asymptotically simplified to yield computationally efficient expressions. Finally, the numeric simulation validating the correctness of the derived results was executed and the practical problem of minimizing capacity reliability for various channel parameters was analyzed.

The remainder of the paper is organized as follows: Section 2 provides some preliminary information about the SIMO system, operating with $\kappa-\mu$ channel model with correlated shadowing, and introduces the analytical definitions of the capacity higher-order statistics; Section 3 states the main contributions of the research: $1)$ the derived analytic representation of the capacity higher-order statistics in terms of the instantaneous signal-to-noise ratio moment generating function, $2)$ the expression for the moment generating function of the instantaneous signal-to-noise ration at the output of the maximum-ratio combiner, $3)$ the resultant analytic and several asymptotic expressions of the CHOS for the SIMO MRC scheme, $4)$ discussion of the derived expression, their comparison and practical suggestions;  Section 4 presents the application of the proposed expressions to the problem of minimum capacity reliability analysis in presence of generalized $\kappa-\mu$ channel with correlated shadowing including several specific subcases (Rician, Rayleigh and Nakagami-$m$); and the conclusions are drawn in Section 5.

\section{Preliminaries and Problem Formulation}
\subsection{$\kappa-\mu$ channel model with correlated shadowing }

Let us assume a communication system with a single transmitting and $N_R$ receiving elements (SIMO case). To achieve maximum gain (both multiplexing and diversity) proper combining of the received signals should be used. For the case when channel state information is available at the receiver maximum-ratio-combing yields the best (optimal) result \cite{Gol05}. For $N_R$-element receive antenna array with the instantaneous signal-to-noise ratio (SNR) per branch $\gamma_i$ the output SNR is defined as \cite{Gol05}:
\begin{equation}
  \gamma=\sum_{i=1}^{N_R}\gamma_i.
\end{equation}

\begin{figure}[!t]
\centerline{\includegraphics[width=\columnwidth]{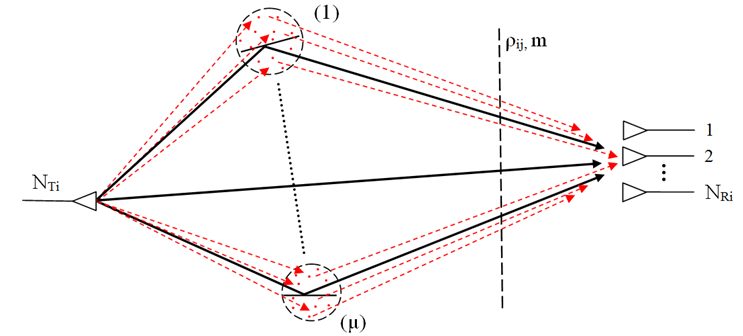}}
\caption{$\kappa \!-\!\mu$ channel with correlated shadowing} \label{fig1}
\end{figure}

Let us describe the propagation subchannels (comprised of the transmitter and each of the receivers) with $\kappa-\mu$ shadowed fading model \cite{Par14} and let the shadowing for all of the subchannels take place with the same intensity $m$, for instance, when the receiving array is compact (hence such a model does not include distributed MIMO (D-MIMO) case), see Fig.~\ref{fig1}.

Assuming the possible correlation of the dominating components shadowing the probability density function of the instantaneous SNR at the receiver is given by \cite{Bha15}:
\begin{eqnarray}\label{pdf}
f_\gamma (\gamma )=\frac{A}{\gamma }\!\!\left(\frac{\eta  \gamma }{\bar{\gamma }} \right)^{\!\!U}\!\!\!e^{-\frac{\eta \gamma }{\bar{\gamma }}}\!\sum _{k=0}^{\infty }\!D_{k} {}_{1} F_{1}  \!\!\left(\!\!mN_R\!+\!k,U;\frac{\eta  \gamma  \lambda }{\bar{\gamma }(1+\lambda )}\!\right)\!,
\end{eqnarray}
with the following set of substitutes:
\begin{equation}
 A=\prod                _{i=1}^{N_R}\left(\frac{\lambda }{\lambda _{i} } \right) ^{m}\!\!\! ,\quad \eta =\sum _{i=1}^{N_R}\mu _{i} (1+\kappa _{i} ), \quad U=\sum _{i=1}^{N_R}\mu _{i},
\end{equation}
where $\bar{\gamma }$ is the average signal-to-noise ratio, i.e. $\bar{\gamma }=\mathbb{E}\left\{\gamma \right\}=2\sigma^{2}\eta $, $\{\mu_1,\ldots \mu_{N_R}\}$ are the numbers of multipath beams (clusters of multipath waves) for each of the receiving antenna elements,  $\{\kappa_1,\ldots \kappa_{N_R}\}$ is the ratio between the total energy of the dominating waves inside each of the clusters and total energy of all other multipath waves for each of the receiving elements, $m$ is the degree of shadowing of the dominating waves and the summation coefficients $D_{k}$ are defined in terms of recurrence relation:
\begin{equation}
\begin{cases}
 \displaystyle D_{k}=\frac{\delta _{k} }{\lambda ^{mN_R+k} \Gamma (U)} \left(1+\frac{1}{\lambda } \right)^{-(mN_R+k)},\\
 \displaystyle \delta _{k}=\frac{m}{k} \sum _{q=1}^{k}\sum _{i=1}^{N_R}\left(1-\frac{\lambda }{\lambda _{i} } \right)^{q} \delta _{k-i},
\end{cases}
\end{equation}
with the initial condition $\delta _{0} =1$; $\lambda $ being the minimum eigenvalue of the eigenspectrum of matrix product ($\mathbf{DC}$) with multipliers
$\mathbf{D}={\rm diag}\left[\frac{\mu _{i} \kappa _{i} }{m} \right]$ (here ${\rm diag}[\cdot]$ stands for the diagonal matrix), and matrix $\mathbf{C}$:
 \begin{eqnarray}
\mathbf{C}=\left[\begin{array}{cccc} {1} & {\sqrt{\rho _{12} } } & {...} & {\sqrt{\rho _{1N_R} } } \\ {\sqrt{\rho _{21} } } & {1} & {...} & {                            \dots } \\ {...} & {...} & {...} & {...} \\ {\sqrt{\rho _{N_R1} } } & {...} & {...} & {1} \end{array}\right],%
\end{eqnarray}
where $\rho_{i,j}$ captures correlation of shadowing processes between $i^{\underline{\rm{th}}}$ and $j^{\underline{\rm{th}}}$ antenna element input signals.

For further analysis a standard exponential model with one-step correlation coefficient $\rho$ would be assumed, i.e. $\rho_{i,j}=\rho^{|i-j|}$ \cite{Sim05}.

\subsection{Capacity higher-order statistics definition}

To describe the variability of channel capacity it is a common practice to introduce so-called capacity higher-order statistics (CHOS) \cite{Sag09}, \cite{Yil12} being the moments of $\log(1+\gamma )$, i.e.

\begin{equation}\label{lambdan}
 \Lambda_n ={\rm {\mathbb E}}\left\{\log ^{n} \left(1+\gamma \right)\right\}=\int _{0}^{\infty }\log ^{n} \left(1+\gamma \right)f_{\gamma } \left(\gamma \right) {\rm d}\gamma.
\end{equation}

Among them the most prominent role (and important from practical point of view) play amount of dispersion $\mathcal{A}o\mathcal{D}$ (AoD) and capacity reliability $\mathcal{R}$ (CR):
\begin{eqnarray}
\displaystyle\mathcal{A}o\mathcal{D}&=&\frac{\mathbb V\{\log(1+\gamma )\}}{\mathbb E\{\log(1+\gamma )\}}=\frac{\Lambda_2}{\Lambda_1}-\Lambda_1,\label{AoD}\\
\displaystyle\mathcal{R}&=&1-\mathcal{A}o\mathcal{D}.\label{R}%
\end{eqnarray}

AoD, being the variance of channel ergodic capacity normalized by its expected value, characterizes the normalized spread of channel capacity (and therefore rate) stochastic variations. It quantifies the distortion in the ergodic capacity per 1-bit information transfer \cite{Yil14} and capacity reliability in its turn is a complementary measure defining its stability.

From a practical perspective, it is important to achieve the maximum possible capacity of the communication link guaranteeing at the same time its minimum distortion. Hence evaluation of AoD and CR is a substantial element of link quality estimation and prediction. To this extent substituting \eqref{pdf} into \eqref{lambdan} to get \eqref{AoD} or \eqref{R} leads to very complex expressions thus an analytical treatment is required.

\section{Fading channel capacity higher-order statistics expressions}

The derivation of closed and asymptotic forms of expressions for $\Lambda_n$, starting with its definition \eqref{lambdan}, can be tackled variously. The one proposed in this research and its main results is stated in the following propositions.

\textit{Proposition 1}: Multipath fading channel capacity higher-order statistics $\Lambda_n$ for arbitrary channel model can be represented in terms of its instantaneous signal-to-noise ratio moment generating function (MGF) $\mathcal{M}_{\gamma }(p)$ as follows:
\begin{equation}\label{Proposition1}
\Lambda_n\!=\!(-1)^n\!\frac{\partial^n}{\partial a^n}\!\left(\!\frac{1}{\Gamma (a)} \int_{0}^{\infty}\!p^{a-1}e^{-p}\mathcal{M}_{\gamma }(-p){\rm d}p\!\right)\!\Biggl|_{a=0}.
\end{equation}
\textit{Proof}: For the proof of \textit{Proposition 1} see \textsc{Appendix~A}.

The obtained expression is much more practical in most cases since for a vast majority of multipath channel statistics used in wireless communication expressions for moment generating functions are generally simpler for further analytical derivations or direct numerical calculations. That is the exact situation for a channel model under consideration. But to employ the obtained result one has to derive MGF for the assumed model, which is stated in \textit{Proposition 2}.

\textit{Proposition 2}: The instantaneous signal-to-noise ratio moment generating function $\mathcal{M}_{\gamma }(p)$ for $N_R$-element SIMO system with the maximum-ratio-combining technique employed at the receiver and $\kappa-\mu$ shadowed fading channel with correlated shadowing can be represented as follows:
\begin{equation}\label{Proposition2}
\mathcal{M}_{\gamma }(p)=C_0 \Gamma (U)\sum _{k=0}^{\infty } \tilde{\delta }_k(\bar{\alpha }-p)^{-m_k}(\alpha-p)^{m_k-U},
\end{equation}
where $\bar{\alpha }=\frac{\alpha }{1+\lambda }$ and $C_0, \bar{\lambda }, m_k$ and $\tilde{\delta }_k$ are defined in \eqref{substitutes}.

\textit{Proof}: For the proof of \textit{Proposition 2} see \textsc{Appendix~B}.

%It is critically to note that the derived representation \eqref{Proposition2} holds for

Combining \textit{Proposition 1} and \textit{Proposition 2} helps to formulate the main result of the research: an analytic ($\Lambda_n^{\kappa-\mu }$) and asymptotic ($\tilde{\Lambda }_n^{\kappa-\mu }$) expressions for the capacity higher-order statistics in presence of multipath $\kappa-\mu$ fading channel with correlated shadowing and MRC SIMO receiving scheme.

\textit{Proposition 3}: In the aforementioned assumptions $\Lambda_n^{\kappa-\mu }$ and $\tilde{\Lambda }_n^{\kappa-\mu }$ are defined as
\begin{equation}\label{Proposition31}
\Lambda_n^{\kappa-\mu }\!\!=\!\frac{(\!-1)^nC_0 \Gamma (U)}{\ln^n(2)}\!\frac{\partial^n}{\partial a^n}\!\left(\sum _{k=0}^{\infty }\! \tilde{\delta }_k \frac{J_a(m_k,\alpha,\lambda,U)}{\Gamma (a)}\!\right)\!\!\Biggl|_{a=0}
\end{equation}
\begin{IEEEeqnarray}{rCl}
\tilde{\Lambda }_n^{\kappa-\mu }&=&\frac{(-1)^nC_0}{\ln^n(2)\alpha^U(1+\lambda )^{-m N_R}} \frac{\partial^n}{\partial a^n}\bigg\{\alpha^a \Gamma (U-a)                               \times                               \IEEEnonumber\\
&&                               \times                               {}\left.\!\left(\sum _{k=0}^{\infty } \bar{\delta }_k\, _2F_1\left(m_k,a;U;\frac{\lambda }{\lambda +1}\right)\right)\right\}\Biggl|_{a=0}\IEEEyesnumber\IEEEyessubnumber\label{Proposition321}\\
\tilde{\Lambda }_n^{\kappa-\mu }&=&\frac{(-1)^nC_0}{\ln^n(2)(1+\lambda )^{-m N_R}}\frac{\partial^n}{\partial a^n}\bigg\{\Psi (U,U+1-a,\alpha )                               \times                               \IEEEnonumber\\
&&                               \times                               \!\left(\sum _{k=0}^{\infty }\!\bar{\delta }_k\!\right)\!-\!\frac{a \lambda \Psi (U,U-a,\alpha )}{\alpha }\! \left(\sum _{k=0}^{\infty } m_k\bar{\delta }_k\!\right)\!+\IEEEnonumber\\
&&+{}\frac{a (a+1) \lambda }{2 \alpha ^2}\Psi (U,U-1-a,\alpha )                               \times                               \IEEEnonumber\\
&&                               \times                               {}\!\left.\left(\sum _{k=0}^{\infty } m_k (\lambda  (m_k+1)+2)\bar{\delta }_k\right)\right\}\Biggl|_{a=0}\IEEEyessubnumber\label{Proposition322}\\
\tilde{\Lambda }_n^{\kappa-\mu }&=&\frac{(\!-1)^nC_0 \Gamma (U)}{\ln^n(2)}\left(\sum _{k=0}^{\infty }\bar{\delta }_k\right)                               \times                               \IEEEnonumber\\
&&                               \times                               {}\frac{\partial^n}{\partial a^n}\Psi (U,-a+U+1,\alpha )\Biggl|_{a=0}.\IEEEyessubnumber\label{Proposition323}%
\end{IEEEeqnarray}
\textit{Proof}: For the proof of \textit{Proposition 3} see \textsc{Appendix~C}.

Up to authors knowledge, \eqref{3C}, i.e. $J_a(m_k,\alpha,\lambda,U)$ has no closed-form solution. Despite it, for practical aspects (for instance for numerical integration) the proposed representation is much simpler than the direct substitution of \eqref{pdf} into \eqref{lambdan}.

At the same time, practical implementation of the derived expression (e.g., for real-time monitoring and prediction of channel capacity fluctuations) requires closed-form expressions; therefore, one has to resort to various forms of approximations \eqref{Proposition321}, \eqref{Proposition322}, \eqref{Proposition323} providing a reasonable error.

The derived expressions \eqref{Proposition321}, \eqref{Proposition322}, \eqref{Proposition323} mainly rely on the asymptotic behaviour of integrand multipliers.  \eqref{Proposition321} is obtained via expansion of the exponent in series and leaving only the zero-order approximation since the dominant impact of the exponent will be in the vicinity of $p=0$. \eqref{Proposition322} is obtained with the expansion of $\left(\frac{\tilde{\alpha }+p}{\alpha+s}\right)^{-m_k}$ in terms of $p$ around $p=0$ and leaving quadratic approximation. On the other hand, \eqref{Proposition323} mainly relies on the fact that for the increasing average SNR the difference between $\alpha$ and $\bar{\alpha }$ will be relatively small; hence, the fraction $\frac{\bar{\alpha }+p}{\alpha+s}$ cancels out.

Direct comparison of the derived expressions demonstrated almost similar approximation quality (in terms of residual error, which does not exceed $1\%$ and in most cases is by several orders of magnitude lower) when all of the parameters $U, \lambda, m_k$ are small. This situation constitutes a case of a small number of antenna elements, poor channel scattering conditions and weakly correlated heavy shadowing. All of the solutions behave well with the increase of $U$. At the same time decrease in shadowing (greater $m$) or increase in its correlation (which results in greater $\lambda$) makes the simplest approximation \eqref{Proposition323} almost inapplicable (since the error increases up to $10\%$ and even more), which is natural as it does not take into account those factors. It can be seen that \eqref{Proposition323} (up to multiplicative factor) is just the zero-order term of \eqref{Proposition322}. On the other hand \eqref{Proposition321} and \eqref{Proposition322} demonstrate excellent approximation quality. At the same time, one has to keep in mind the recursion procedure in index $k$ which in practice has to be terminated at some point and generally implies the strongest restrictions on computational time and complexity. Because of this factor, it should be noted that recursions in \eqref{Proposition322} and \eqref{Proposition323} do not require evaluation of complex mathematical structures (like hypergeometric function as in \eqref{Proposition321} and in the initial equation, although there are many mathematical libraries, which provide numerically satisfactory procedures computing them).

From a practical point of view, the channel estimation procedure, which is usually done beforehand, yields channel parameters estimates (i.e. $m$, $\mu$, $\kappa$, $\rho$) that can serve as a guide for the choice of approximation (\eqref{Proposition321}, \eqref{Proposition322} or \eqref{Proposition323}).

Although the derived results are mainly used herein for further AoD and CR calculation, they are generally not limited to those factors and can be applied to any channel quality metrics that exploit capacity higher-order statistics \eqref{lambdan}.

\section{Practical implementation to the problem of maximum capacity reliability analysis. Simulation and Results}
\subsection{Practical application: maximum capacity reliability analysis}

To illustrate the application of the obtained results, the problem of maximum  AoD (or minimum CR) analysis is assumed.

\begin{figure}[h]
\centerline{\includegraphics[width=\columnwidth]{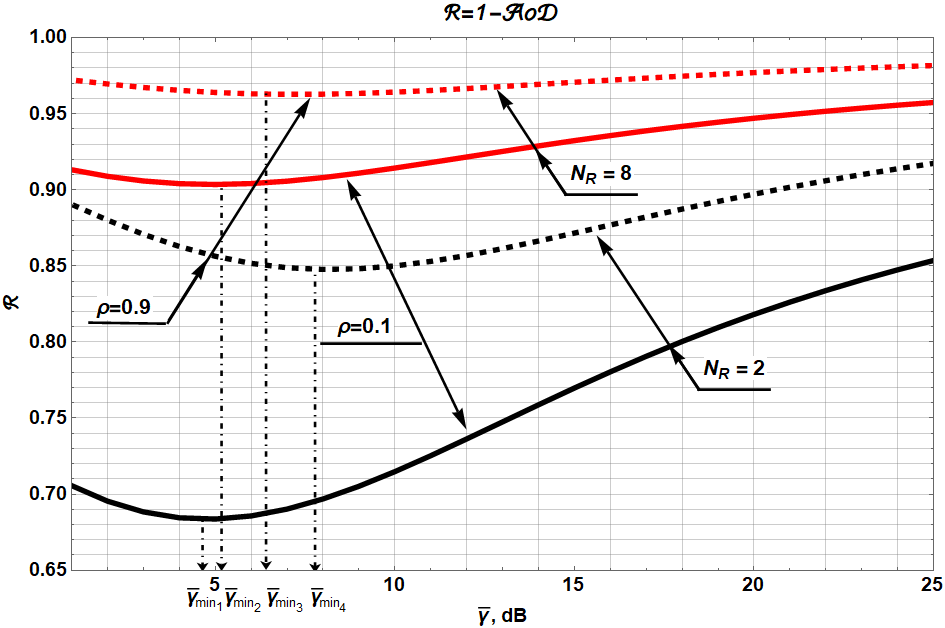}}
\caption{$\mathcal{R}$ for $\kappa \!-\!\mu$-correlated shadowed channel model for two distinct configurations }
\label{fig2}
\end{figure}

It is a common fact that in many practical situations (specific range of SNR) AoD/CR can be considered as single extremum curves with respect to SNR \cite{Yil12}, \cite{Yil14}. At the same time the absolute value of $\mathcal{R}_{\min}=\min(\mathcal{R})$ (or $\mathcal{A}o\mathcal{D}_{\max}=\max(\mathcal{A}o\mathcal{D})$) and the SNR at which it is attained, i.e. $\gamma_{\mathcal{R}_{\min}}=\gamma_{\mathcal{A}o\mathcal{D}_{\max}}=\operatorname*{arg\,min}_{\gamma } \mathcal{R}(\gamma )$ heavily depends on the channel and communication system parameters (for instance, see Fig.~\eqref{fig2}).

Since, as it was mentioned earlier, the assumed model incorporates a wide variety of less general channel models, the following cases were analysed:
\begin{itemize}
  \item Rician fading with correlated shadowing($\mu=1$, $m\to \infty$):  one multipath cluster of waves with a dominant component exhibiting no shadowing, with varying Rician K-factor and $\rho$.
  \item  Nakagami-$m$ fading with correlated shadowing($\kappa \to 0, m\to \infty$):  two multipath clusters of waves with no dominant component, variable Nakagami-$m$ parameter and $\rho$.
  \item  Rayleigh fading with correlated shadowing($\kappa \to 0, \mu=1, m\to \infty$): a single cluster of multipath waves with no dominant component and variable $\rho$.
  \item  Generalised case ($\kappa-\mu$ with correlated shadowing) with variable $\kappa, \mu, m$ and $\rho$.
\end{itemize}

\subsection{Simulation and Results}
\begin{figure}[t]
\centerline{\includegraphics[width=\columnwidth]{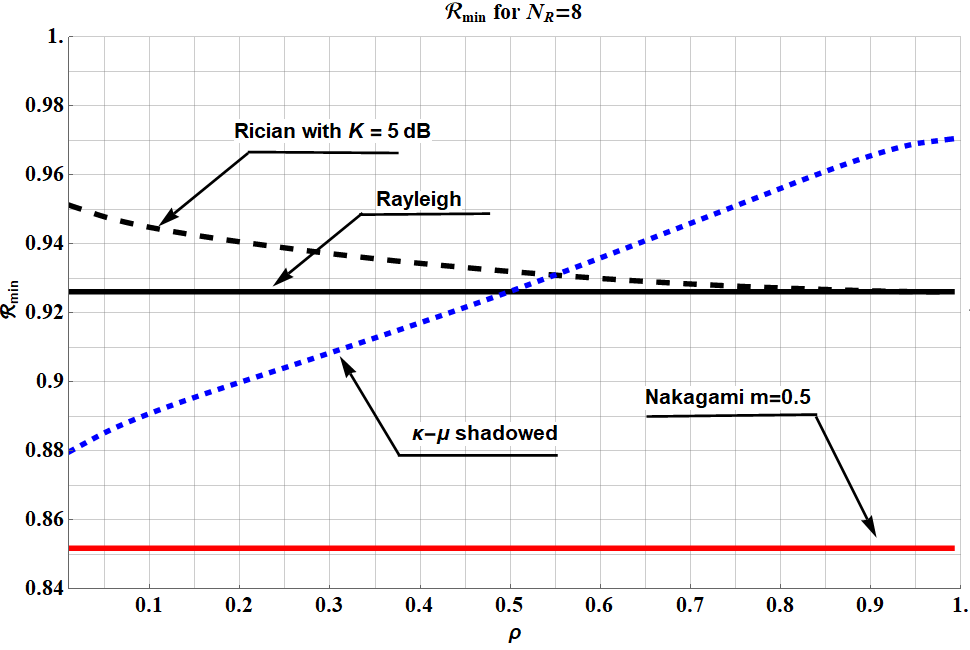}}
    \caption{ Model cross-comparison for $\mathcal{R}_{\min}$ and varying correlation}
\label{fig3}
\end{figure}

\begin{figure}[t]
\centerline{\includegraphics[width=\columnwidth]{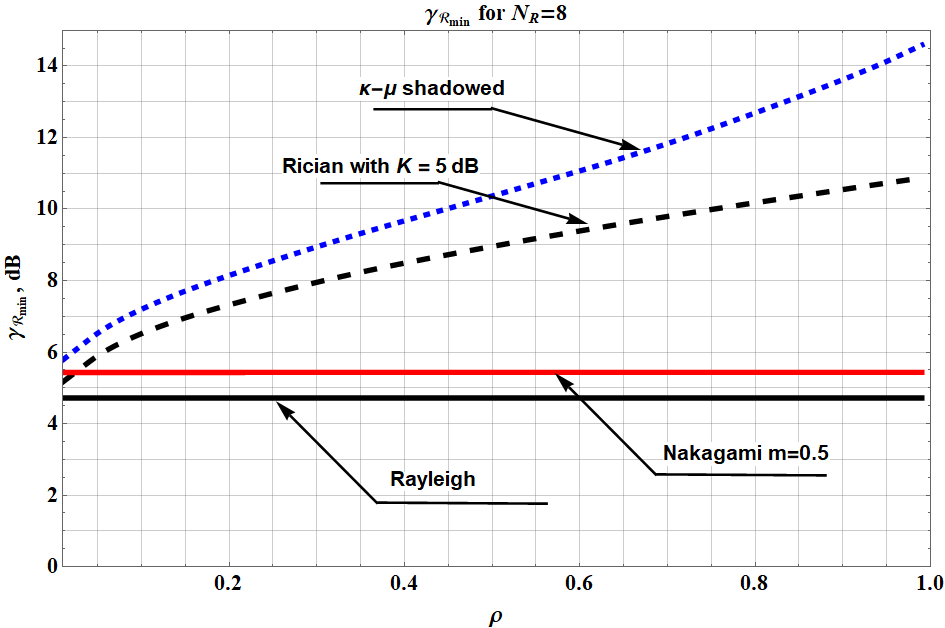}}
     \caption{Model cross-comparison for $\gamma_{\mathcal{R}_{\min}}$ and varying correlation }
\label{fig4}
\end{figure}

Analysing the impact of shadowing correlations (see Fig.~\eqref{fig3}--\eqref{fig4}), one can see that $\mathcal{R}_{\min}$ as well as $\gamma_{\mathcal{R}_{\min}}$ exhibits a strong dependence on $\rho$ for two out of four models, which is natural, since for Rayleigh and Nakagami-$m$ cases $\kappa \to 0$, hence the impact of dominant components' correlation is absent.

For $\kappa \!-\!\mu$ model numeric simulations demonstrated the high sensitivity of $\mathcal{R}_{\min}$ (and low sensitivity of $\gamma_{\mathcal{R}_{\min}}$) to the number of elements in case of small shadowing correlation, moreover, all of the analysed effects become more pronounced with the decrease of the one-step correlation coefficient.

It should be noted that the obtained results are not limited to those models; hence, the performed research helps to get a better insight into the channel propagation effects.

Cross-comparison of the simulation results demonstrated that for the Rician case, the increase in $\rho$ decreases CR to the value of CR for the Rayleigh case, which means that a higher correlation of shadowing of a single dominant component almost diminishes its impact. Contrary to that, for the generalised case (for example, with $\mu=2.5$ and $\kappa=10$~dB as in Fig.~\eqref{fig3}-\eqref{fig4}) the increase in $\rho$ improves CR, intersecting CR level for the Rayleigh case at $\rho=0.5$.

Although one can see a clear improvement in the channel capacity stability (for the generalised case) with the increase in $\rho$, it is achieved at higher $\gamma_{\mathcal{R}_{\min}}$ (the curve $\mathcal{R}(\bar{\gamma })$ shifts to the right, see Fig.~\eqref{fig2}), which should not be the case for practical applications, since it makes the communication rate fluctuations less predictable at higher SNR.


Quantifying the impact of correlation effects and using the criteria proposed in \cite{Dur99} (i.e. the correlation is assumed to be sufficient if $\rho \geq 0.4$), one can see that the $\kappa-\mu$ model is about $0.07$ bit/s/Hz more efficient than Nakagami-$m$ (in terms of $\mathcal{R}_{\min}$) and loses about $0.01$ bit/s/Hz and $0.02$ bit/s/Hz to Rayleigh and Rician (with $K=5$~dB) model respectively. It should be noted that the improvement in minimum CR for the $\kappa-\mu$ at the high shadowing correlations is obtained in exchange for the rapid decrease of the reception diversity \cite{Pau08}, which is a highly undesirable situation.

\subsection{Discussions}
It should be noted that the problem of HoS analysis in the presence of $\kappa-\mu$ correlated shadowed fading has been addressed earlier in \cite{Zha17} and \cite{Gvo19, Gvo21}. But, as it was pointed out, the main goal of the research was to establish a connection between the HoS analysis and the MGF approach, which is the primary result of the paper. Moreover, this helped to derive closed-form and approximating expressions. This differs the proposed research from \cite{Zha17}, where the classical PDF methodology was exploited. Furthermore, the approximations given herein are computationally less complex, i.e. one needs only confluent hypergeometric functions, which are readily implemented in almost all computational software and have many numerically efficient representations, instead of sums of Meijer G-functions (which are generally much more complex for implementation); and the derived expressions use only a single summation, rather than double (compare \eqref{Proposition321}-\eqref{Proposition323} with equations  (40)-(41) in \cite{Zha17}). On the other hand, in \cite{Gvo19} only qualitative simulation (i.e. no analytic results) was performed, and in  \cite{Gvo21}, even though that the similar MGF approach was chosen (but in a formulation based on the results from \cite{Yil14}), no closed-form solution or approximations were derived.  In fact, non of the researchers addressed the problem of maximum capacity reliability analysis, which was performed herein.

\section{Conclusion}
The proposed research considers an analytical treatment of wireless channel capacity higher-order statistics for diversity reception in the presence of composite fading and correlated shadowing. For a SIMO system employed with a maximum-ratio combining scheme at the receiving end an analytic expression for the capacity higher-order statistics is derived in terms of the moment generating function of the instantaneous output signal-to-noise ratio. The asymptotic approximations of the proposed solution are evaluated and analyzed. The performed derivations are exemplified by the problem of the  capacity reliability minimizing for various fading channel conditions. It is demonstrated that for the generalized $\kappa-\mu$ fading channel (with $\kappa=10$~dB and $\mu=2.5$) and correlated shadowing of the dominating components the minimum value of the CR (due to rate-distortion as a result of channel instantaneous fluctuations) is attained at the higher SNR rather than for Rayleigh, Rician or Nakagami-$m$ channel models for any one-step correlation coefficient. Moreover, the absolute of the CR for the $\kappa-\mu$ fading model coincides with that of the Rayleigh model for $\rho=0.5$.

\section*{Appendix A \\ Proof of Proposition 1}
First, to ease the notation, let us for convenience rewrite \eqref{pdf} with the following set of substitutes:
\begin{IEEEeqnarray}{C}
 f_{\gamma }(\gamma )=C_0 e^{-\alpha  \gamma } \gamma ^{U-1} \sum _{k=0}^{\infty } \tilde{\delta }_k \, _1F_1\left(m_k;U;\alpha  \gamma  \bar{\lambda }\right),\label{pdf_new}\\
 \begin{cases}\label{substitutes}
 \displaystyle C_0=\frac{A\alpha^U (1+\lambda )^{-m N_R} m}{\ln^n(2)\Gamma (U)}, \qquad \bar{\lambda }=\frac{\lambda }{1+\lambda }\\ %\qquad \tilde{\delta }_k=\frac{\delta _k}{m(1+\lambda )^k}
 \displaystyle m_k=m N_R+k, \qquad \qquad\alpha=\frac{\eta }{\bar{\gamma }},
 \end{cases}
\end{IEEEeqnarray}
where $C_0$ is the multiplicative factor that does not depend upon the summation index $k$ and instantaneous SNR, $\Gamma (         \cdot )$ is the Euler gamma function and new recursion coefficients $ \tilde{\delta }_k$.
\begin{equation*}
   \tilde{\delta }_k=\frac{1}{k(1+\lambda )^k}\sum_{q=1}^{k}g_q  \tilde{\delta }_{k-q}, \qquad g_q=\sum _{i=1}^{N_R}\left(1-\frac{\lambda }{\lambda _{i} }\right)^{q}.
\end{equation*}

Sequential application of the $n$-th power logarithm standard representation and Schwinger parametrization yield:
\begin{eqnarray*}
  \ln^n(1+\gamma )\!\!\!&=&\!\!\!(-1)^n\frac{\partial^n}{\partial a^n}(1+\gamma )^{-a}\Biggl|_{a=0} \\
   \!\!\!&=&\!\!\!(-1)^n\frac{\partial^n}{\partial a^n}\left(\frac{1}{\Gamma (a)}\int_{0}^{\infty}e^{-p(1+\gamma )}p^{a-1}{\rm d}p\right) \Biggl|_{a=0}.
\end{eqnarray*}

Hence the higher-order statistics can be represented as follows:
\begin{equation*}
  \Lambda_n\!=\!(-1)^n\frac{\partial^n}{\partial a^n}\!\left(\!\frac{1}{\Gamma (a)} \!\int_{0}^{\infty}\!\!p^{a\!-\!1}e^{-p}\!\underbrace{\int_{0}^{\infty}\!\!e^{-p\gamma }\!f_{\gamma } (\gamma ){\rm d}\gamma }_{\mathcal{M}_{\gamma }(-p)}{\rm d}p\right)\!\Biggl|_{a=0},
\end{equation*}
which completes the proof.

\section*{Appendix B \\ Proof of Proposition 2}
Combining the definition of the moment generating function $\mathcal{M}_{\gamma }(p)$ with \eqref{pdf_new} and \eqref{substitutes} yields:
\begin{eqnarray*}
  \mathcal{M}_{\gamma }(p)\!\!\!&=&\!\!\!\!C_0 \sum _{k=0}^{\infty } \tilde{\delta }_k \int_{0}^{\infty}\!\gamma ^{U-1} \!\, _1F_1\left(m_k;U;\alpha \gamma  \bar{\lambda }\right)e^{-(\alpha-p)\gamma }{\rm d}\gamma \\
  \!\!\!\!&=&\!\!\!\!C_0 \sum_{k=0}^{\infty } \tilde{\delta }_k \mathfrak{M}_{U}\left\{\, _1F_1\left(m_k;U;\alpha \gamma  \bar{\lambda }\right)e^{-(\alpha-p)\gamma }\right\},
\end{eqnarray*}
where $\mathfrak{M}_{U}(               \cdot )$ defines Mellin transform evaluated at $U$. Applying equation (6.9.9) from \cite{Bat54} one gets
\begin{IEEEeqnarray}{l}
\hspace{-30pt}\mathcal{M}_{\gamma }(p)=C_0 \sum _{k=0}^{\infty } \tilde{\delta }_k \frac{\Gamma (U)}{(\alpha-p)^{U}}\!\, _2F_1\left(\!m_k,U;U;\frac{\alpha \bar{\lambda }}{\alpha-p}\!\right)\IEEEyessubnumber\label{2B}\\
\hspace{3pt}=C_0 \Gamma (U)\! \sum _{k=0}^{\infty } \tilde{\delta }_k \!\!\left(1-\frac{\alpha \bar{\lambda }}{\alpha-p}\right)^{-m_k}\!\!\!(\alpha-p)^{-U},\IEEEyessubnumber%
\end{IEEEeqnarray}
here $\, _2F_1\left(\cdot,\cdot;\cdot;                                \cdot \right)$ is the Gauss hypergeometric function and the last line was obtained with the help of \cite{DLMF} (equation 15.4.6), i.e. $\, _2F_1\left(a,b,b,z\right)=(1-z)^{-a}$.
%=C_0 \sum _{k=0}^{\infty }\!\tilde{\delta }_k\mathscr{M}_{U}\!\left\{\!\, _1F_1\!\left(m_k;U;\alpha \gamma  \bar{\lambda }\right)\!e^{-(\alpha-p)\gamma }\!\right\}\IEEEyesnumber\IEEEyessubnumber\\ \hspace{3pt}
One can notice that the intermediate result \eqref{2B} is equivalent to the one obtained in \cite{Bha15} (up to the sign of $p$), but the proposed herein derivation is sufficiently simpler and does not require such a heavyweight approach as Meijer-G function transformation.
Reducing the inner terms completes the proof.

\section*{Appendix C \\ Proof of Proposition 3}
As for the first part of the proposition, equation \eqref{Proposition31} is a simple extent of \textit{Proposition 1} and \textit{Proposition 2} with \eqref{Proposition2} being substituted into \eqref{Proposition1}. After rearranging multipliers one gets the expression for $\Lambda_n^{\kappa-\mu }$ in terms of integral $J_a(m_k,\alpha,\lambda,U)$:
\begin{equation}\label{3C}
  J_a(m_k,\alpha,\lambda,U)\!=\!\int_0^{\infty }\!\!\! p^{a-1}\! (\alpha +p)^{-u}\! \left(\!\frac{\bar{\alpha }+p}{\alpha +p}\!\right)^{\!-m}e^{-p}{\rm d}p.
\end{equation}

Noticing that the dominant impact of the integrand is reached in the vicinity of $p=0$ one can omit exponential term (or equivalently using its zero-order Taylor approximation), hence (see \cite{Gra96}):
\begin{IEEEeqnarray}{rCl}
&&\hspace{-1cm}J_a(m_k,\alpha,\lambda,U)                              \simeq                              \int_0^{\infty }\!\!\! p^{a-1}\! (\alpha +p)^{-u}\! \left(\!\frac{\bar{\alpha }+p}{\alpha +p}\!\right)^{\!-m}{\rm d}p\IEEEyesnumber\IEEEyessubnumber\\
&&\hspace{0cm}=\frac{\alpha^{a-U}\Gamma (a)\Gamma (U-a)}{(1+\lambda )^{-m_k}\Gamma (U)}\,_2F_1\left(m_k,a;U;\frac{\lambda }{1+\lambda }\right).\IEEEyessubnumber%
\end{IEEEeqnarray}

After cancelling gamma functions one gets \eqref{Proposition321}, where $\bar{\delta }$ is defined through the similar recursion procedure as earlier, but without $(1+\lambda )^{-k}$ factor, i.e. $\bar{\delta }_k=\sum_{q=1}^{\infty}g_q  \bar{\delta }_{k-q}$.

Using the same reasoning as earlier but expanding $\left(\frac{\bar{\alpha }+p}{\alpha +p}\right)^{\!-m}$ in Taylor series (up to the second order) yields:
\begin{IEEEeqnarray}{rCl}
\hspace{-1cm}J_a(m_k,\alpha,\lambda,U)&                              \simeq                              &(1+\lambda )^{m_k} \int_0^{\infty }\!\!\! p^{a-1}\! \left\{1-\frac{\lambda  m p}{\alpha }+\right.\IEEEnonumber\\
&&\hspace{-2cm}+\left.\frac{\lambda  m p^2 (\lambda +\lambda  m+2)}{2 \alpha ^2}+O\left(p^3\right)\right\}(\alpha +p)^{-u}\! e^{-p}{\rm d}p.\IEEEyesnumber%
\end{IEEEeqnarray}

Sequential termswise integration (omitted here for the sake of brevity) delivers \eqref{Proposition322}, where $\Psi (\cdot,\cdot,                              \cdot )$ is the Tricomi confluent hypergeometric function \cite{DLMF}.

Observing that for small $\lambda$ the difference between $\bar{\alpha }=\frac{\alpha }{1+\lambda }$ and $\alpha $ is negligible one can use a zero-order approximation of $\left(\frac{\bar{\alpha }+p}{\alpha +p}\right)^{\!-m}$, hence obtaining:
\begin{IEEEeqnarray}{rCl}
&&\hspace{-1cm}J_a(m_k,\alpha,\lambda,U)                              \simeq                              \int_0^{\infty }\!\!\! p^{a-1}\! (\alpha +p)^{-u}\! e^{-p}{\rm d}p\IEEEyesnumber\IEEEyessubnumber\\
&&\hspace{0cm}=\Gamma (U)\Psi (U,a-a+1,\alpha ).\IEEEyessubnumber%
\end{IEEEeqnarray}

Hence \eqref{Proposition323} follows.

%%%%%%%%%%%%%%%%%%%%%%%%%%%%%%%%%%%%%%%%%%%%%%%%%%%%%%%%%%%%%%%%%%%%%%%%%5
%\section{Section title}
%\label{sec:1}
%Text with citations \cite{RefB} and \cite{RefJ}.
%\subsection{Subsection title}
%\label{sec:2}
%as required. Don't forget to give each section
%and subsection a unique label (see Sect.~\ref{sec:1}).
%\paragraph{Paragraph headings} Use paragraph headings as needed.
%\begin{equation}
%\end{equation}

%\begin{acknowledgements}
%\end{acknowledgements}

\bibliographystyle{IEEEtran}
\bibliography{IEEEabrv,Bibliography}

\begin{thebibliography}{10}
\providecommand{\url}[1]{#1}
\csname url@samestyle\endcsname
\providecommand{\newblock}{\relax}
\providecommand{\bibinfo}[2]{#2}
\providecommand{\BIBentrySTDinterwordspacing}{\spaceskip=0pt\relax}
\providecommand{\BIBentryALTinterwordstretchfactor}{4}
\providecommand{\BIBentryALTinterwordspacing}{\spaceskip=\fontdimen2\font plus
\BIBentryALTinterwordstretchfactor\fontdimen3\font minus
  \fontdimen4\font\relax}
\providecommand{\BIBforeignlanguage}[2]{{%
\expandafter\ifx\csname l@#1\endcsname\relax
\typeout{** WARNING: IEEEtran.bst: No hyphenation pattern has been}%
\typeout{** loaded for the language `#1'. Using the pattern for}%
\typeout{** the default language instead.}%
\else
\language=\csname l@#1\endcsname
\fi
#2}}
\providecommand{\BIBdecl}{\relax}
\BIBdecl

\bibitem{Pau08}
A.~Paulraj, R.~Nabar, and D.~Gore, \emph{Introduction to space-time wireless
  communications}.\hskip 1em plus 0.5em minus 0.4em\relax Cambridge University
  Press, 2008.

\bibitem{Gol05}
A.~Goldsmith, \emph{Wireless communications}.\hskip 1em plus 0.5em minus
  0.4em\relax Cambridge University Press, 2005.

\bibitem{Sha20}
K.~Shafique, B.~A. Khawaja, F.~Sabir, S.~Qazi, and M.~Mustaqim, ``Internet of
  things ({IoT}) for next-generation smart systems: A review of current
  challenges, future trends and prospects for emerging 5{G}-{I}o{T}
  scenarios,'' \emph{{IEEE} Access}, vol.~8, pp. 23\,022--23\,040, 2020.

\bibitem{Agi16}
M.~Agiwal, A.~Roy, and N.~Saxena, ``Next generation 5{G} wireless networks: A
  comprehensive survey,'' \emph{{IEEE} Communications Surveys {\&} Tutorials},
  vol.~18, no.~3, pp. 1617--1655, 2016.

\bibitem{Kho20}
E.~Khorov, I.~Levitsky, and I.~F. Akyildiz, ``Current status and directions of
  {IEEE} 802.11be, the future {W}i-{F}i 7,'' \emph{{IEEE} Access}, vol.~8, pp.
  88\,664--88\,688, 2020.

\bibitem{Den20}
C.~Deng, X.~Fang, X.~Han, X.~Wang, L.~Yan, R.~He, Y.~Long, and Y.~Guo, ``{IEEE}
  802.11be {W}i-{F}i 7: New challenges and opportunities,'' \emph{{IEEE}
  Communications Surveys {\&} Tutorials}, vol.~22, no.~4, pp. 2136--2166, 2020.

\bibitem{Sim05}
M.~K. Simon and M.-S. Alouini, \emph{Digital Communications Over Fading
  Channels}, 2nd~ed.\hskip 1em plus 0.5em minus 0.4em\relax John Wiley \& Sons,
  Inc., 2005.

\bibitem{Yac07a}
M.~D. Yacoub, ``The $\kappa$-$\mu$ distribution and the $\eta$-$\mu$
  distribution,'' \emph{IEEE Antennas and Propagation Magazine}, vol.~49,
  no.~1, pp. 68--81, 2007.

\bibitem{Yac07b}
------, ``The $\alpha$-$\mu$ distribution: A physical fading model for the
  stacy distribution,'' \emph{IEEE Transactions on Vehicular Technology},
  vol.~56, no.~1, pp. 27--34, 2007.

\bibitem{Par14}
J.~F. Paris, ``Statistical characterization of $\kappa$-$\mu$ shadowed
  fading,'' \emph{IEEE Transactions on Vehicular Technology}, vol.~63, no.~2,
  pp. 518--526, 2014.

\bibitem{Bha15}
M.~R. {Bhatnagar}, ``On the sum of correlated squared $\kappa-\mu$ shadowed
  random variables and its application to performance analysis of {MRC},''
  \emph{IEEE Transactions on Vehicular Technology}, vol.~64, no.~6, pp.
  2678--2684, June 2015.

\bibitem{Sag09}
N.~C. Sagias, F.~I. Lazarakis, A.~A. Alexandridis, K.~P. Dangakis, and G.~S.
  Tombras, ``Higher order capacity statistics of diversity receivers,''
  \emph{Wireless Personal Communications}, vol.~56, no.~4, pp. 649--668, 2009.

\bibitem{Yil12}
F.~Yilmaz and M.-S. Alouini, ``On the computation of the higher-order
  statistics of the channel capacity over generalized fading channels,''
  \emph{IEEE Wireless Communications Letters}, vol.~1, no.~6, pp. 573--576, dec
  2012.

\bibitem{Yil14}
F.~Yilmaz, H.~Tabassum, and M.-S. Alouini, ``On the computation of the higher
  order statistics of the channel capacity for amplify-and-forward multihop
  transmission,'' \emph{IEEE Transactions on Vehicular Technology}, vol.~63,
  no.~1, pp. 489--494, jan 2014.

\bibitem{Tsi16}
T.~A. Tsiftsis, F.~Foukalas, G.~K. Karagiannidis, and T.~Khattab, ``On the
  higher order statistics of the channel capacity in dispersed spectrum
  cognitive radio systems over generalized fading channels,'' \emph{{IEEE}
  Transactions on Vehicular Technology}, vol.~65, no.~5, pp. 3818--3823, may
  2016.

\bibitem{Pep18}
K.~P. Peppas, P.~T. Mathiopoulos, J.~Yang, C.~Zhang, and I.~Sasase,
  ``High-order statistics for the channel capacity of {EGC} receivers over
  generalized fading channels,'' \emph{{IEEE} Communications Letters}, vol.~22,
  no.~8, pp. 1740--1743, aug 2018.

\bibitem{Dur99}
G.~D. Durgin and T.~S. Rappaport, ``Effects of multipath angular spread on the
  spatial cross-correlation of received voltage envelopes,'' in \emph{1999 IEEE
  49th Vehicular Technology Conference}, vol.~2.\hskip 1em plus 0.5em minus
  0.4em\relax {IEEE}, pp. 996--1000.

\bibitem{Zha17}
J.~Zhang, X.~Chen, K.~P. Peppas, X.~Li, and Y.~Liu, ``On high-order capacity
  statistics of spectrum aggregation systems over $\kappa$- $\mu$ and $\kappa$-
  $\mu$ shadowed fading channels,'' \emph{IEEE Transactions on Communications},
  vol.~65, no.~2, pp. 935--944, feb 2017.

\bibitem{Gvo19}
A.~Gvozdarev, P.~E. Patralov, and I.~V. Kanaev, ``An analysis of ergodic
  capacity higher-order statistics for multiantenna communication system in
  presence of generalized shadowed fading channel,'' in \emph{2019 Systems of
  Signal Synchronization, Generating and Processing in Telecommunications
  (SYNCHROINFO)}.\hskip 1em plus 0.5em minus 0.4em\relax {IEEE}, jul 2019.

\bibitem{Gvo21}
A.~Gvozdarev and P.~Patralov, ``Probabilistic analysis of generalised statistic
  model for multipath channel of {SIMO} systems with fading and correlated
  shadowing,'' \emph{Informatics and Automation}, vol.~20, no.~3, pp. 727--749,
  may 2021.

\bibitem{Bat54}
H.~Bateman and A.~Erd{'e}lyi, \emph{Tables of integral transforms}.\hskip 1em
  plus 0.5em minus 0.4em\relax McGraw-Hill, 1954.

\bibitem{DLMF}
F.~W.~J. Olver, \emph{NIST handbook of mathematical functions}.\hskip 1em plus
  0.5em minus 0.4em\relax Cambridge University Press, 2010.

\bibitem{Gra96}
I.~S. Gradshteyn, A.~Jeffrey, and I.~M. Ryzhik, \emph{Table of integrals,
  series, and products}.\hskip 1em plus 0.5em minus 0.4em\relax Academic Press,
  1996.

\end{thebibliography}

\end{document}